\documentclass[12pt]{article}      
\usepackage{graphics}
\usepackage{amsthm}
\usepackage{amssymb}
\usepackage{amsmath}
\usepackage{amsfonts}
\usepackage{epic}
\usepackage{hyperref}
\usepackage{epsfig}
\usepackage{bm}
\usepackage{amscd}

\theoremstyle{plain}

\theoremstyle{definition}

\def\be{\begin{equation}}
\def\ee{\end{equation}} 
\title{The String Theory on AdS$_3$ as a Marginal Deformation of a Linear Dilaton Background}

\smallskip

\author{Gaston Giribet${}^{1,2}$ and Yu Nakayama${}^3$}

\begin{document}
\begin{titlepage}
\begin{flushright}
\end{flushright}
\begin{center}
\noindent{{\LARGE{The String Theory on AdS$_3$ as a Marginal Deformation of a Linear Dilaton Background}}} 

\smallskip
\smallskip
\noindent{\large{Gaston Giribet}}
\end{center}
\smallskip
\smallskip
\centerline{Department of Physics, Universidad de Buenos Aires, FCEN,}
\centerline{{\it Ciudad Universitaria, 1428. Buenos Aires, Argentina}}
\smallskip
\centerline{Department of Physics, Universidad Nacional de La Plata, IFLP,}
\centerline{{\it C.C. 67, 1900, La Plata, Argentina}}
\smallskip

\smallskip

\begin{abstract}
We investigate $N$-point string scattering amplitudes in $AdS_3$ space. Based on recent observations on the solutions of KZ and BPZ-type differential equations, we discuss how to describe the string theory in $AdS_3$ as a marginal deformation of a (flat) linear dilaton background. This representation resembles the called ``discrete light-cone Liouville'' realization as well as the FZZ dual description in terms of the sine-Liouville field theory. Consequently, the connection and differences between those and this realization are discussed.  The free field representation presented here permits to understand the relation between correlators in both Liouville and WZNW theories in a very simple way. Within this framework, we discuss the spectrum and interactions of strings in Lorentzian $AdS_3$.

\end{abstract}

\end{titlepage}



\newpage


\section{Introduction}
        
In this paper we study a representation of the $N$-point correlation functions describing string
scattering amplitudes in Euclidean $AdS_3$. The tree-level amplitudes of
strings in this space were extensively studied in the literature and
received renewed attention since the formulation of the $AdS/CFT$
correspondence. The first exact computation of three and two-point functions
was done in Ref. \cite{B,BB} and was subsequently extended and studied in
rigorous detail in Ref. \cite{T1,T2,T3} by Teschner. The interaction processes of
winding string states were studied latter \cite{GN3,MO3}, after
Maldacena and Ooguri proposed the inclusion of those states in the spectrum of
the theory \cite{MO1}. Besides, several formalisms yielding consistent results were
shown to be useful tools for studying the correlators in this non-compact CFT
\cite{Satoh,GK1,GK2,HS,HOS,IOS}; however, the most fruitful
method for analyzing the functional properties of these observables was, so
far, the employing of the analogies existing between this CFT and its
relative: the Liouville field theory. Though our understanding of correlation
functions in Euclidean and Lorentzian $AdS_3$ was substantially increasing in
the last years, several features remain still as open questions: The
factorization properties of the generic four-point function and the closeness
of the operator product expansion of unitary states are, perhaps, the most
important puzzles from the viewpoint of the applications to string theory. The
motivation of this work is that of studying some aspects of the $N$-point
functions in this CFT for generic $N$. The general expression for the case $N>3$ is not known, but a
new insight about its functional form appeared recently due to the discovery
of a very direct relation between these and analogous correlators in Liouville
field theory \cite{S,RT,R,GN}. Although the general expression of the $N$-point function is
not known even for the Liouville field theory, such a connection between correlators in both theories still
represents a great opportunity for exploring some aspects of their analytic
properties. We will show here how the existence of such a
connection leads to find a way of representing the worldsheet theory of $AdS_3
$ strings in terms of a tachyonic linear dilaton background. This is an important observation due to the fact that, precisely, the representation of the $SL(2,R)_k/U(1)$ WZNW in terms of a tachyonic background (that is believed to be dual of it) turned out to be a useful method for describing the physics of strings in the black hole background; see for instance \cite{GK1,GK2,Takayanagi1,HK}. Among those results obtained in the last years by means of the employing of such connection between curved and linear-dilaton tachyonic backgrounds, the one we find most important is the construction of a matrix model for the 2D black hole geometry \cite{KKK, KMS} which clearly shows the relevance that the study of this kind of duality has within the context of string theory. Here, we are interested in establishing a precise connection between string theory in $AdS_3$ and a similar tachyonic linear dilaton background; and we describe it in the next section. To be precise, the paper is organized as follows: In section 2 we discuss how the Ribault formula, connecting correlators in both WZNW and Liouville CFTs, leads to a free field realization of the worldsheet theory on $AdS_3$. Then, it turns out that this realization is a Liouville-like model on the worldsheet and can be thought of as a dual description as well. We consequently analyze the relation with other approaches and the CFT formulation carefully. In section 3 we comment on the spectrum and interactions of strings in Lorentzian $AdS_3$ and make some remarks on the integral representation of the free field formulation we propose. In particular, we study the factorization limit of four-point functions in both WZNW and sine-Liouville model and focus our attention on the scattering processes that violate the winding number conservation in $AdS_3$. We show that intermediate states with both winding number $\omega =0$ and $\omega =1$ arise in the internal channels of four-point amplitudes. We also discuss how the upper bound for the violation of total winding conservation in $N$-point correlators naturally appears in this context.

\section{The linear dilaton background}

\subsection{The Ribault-Teschner formula}

Recently, a new formula connecting correlators in both $SL(2,C)_k /SU(2)$ WZNW
and Liouville conformal theories was discovered by Teschner and Ribault \cite
{RT}, who studied an improved version of a previous result due to Stoyanovsky
\cite{S}. In Ref. \cite{R}, Ribault extended the result of
\cite{RT} by achieving to write down a formula connecting the $N$-point
correlation functions in Euclidean $AdS_3$ string theory and certain subset of
$M$-point functions in Liouville field theory, where the relation between $N$
and $M$ turns out to be determined by the winding number of the string states
involved in the correlators. If $\Phi _{j,m,\bar m}^{\omega } (z)$ represents
the vertex operators in the WZNW model creating string states belonging to
the flowed representations of $\hat {sl(2)}_k \times \hat{sl(2)}_k$ affine algebra, and $V_
{\alpha } (z)$ represents the vertex operators of the quantum Liouville theory
with cosmological term $\mu e^{\sqrt {2} b \varphi (z)}$, then the Ribault
formula reads
\begin{eqnarray}
\langle \prod _{i=1}^N \Phi _{j_i,m _i,\bar {m}_i}(z_i)  \rangle _{SL(2)} &=& N_k (j_1, ...j_N;m_1, ...m_N) \prod_{r=1}^{M} \int d^2w_r \ F_k (z_1, ...z_N ; w_1, ... w_{M}) \times \nonumber \\ && \times \langle \prod _{t=1}^{N} V_{\alpha _t} (z_t)   \prod _{r=1}^M V_{-\frac {1}{2b}}(w_r)  \rangle _L \ , \label{rt}
\end{eqnarray}
where the normalization factor is given by 
\begin{eqnarray}
N_k (j_1, ...j_N;m_1, ...m_N) &=&  \frac {2\pi ^{3-2N}b }{M! \ c_k^{M+2}} (\pi ^2 \mu b^{-2})^{-s} \prod _{i=1}^{N} \frac {c_k \ \Gamma (-m_i-j_i)}{\Gamma (1+j_i+\bar {m}_i)} \label{N}
\end{eqnarray}
while the $z$-dependent function is
\begin{eqnarray}
F_k (z_1, ...z_N ; w_{1}, ... w_{M}) &=& \frac {  \prod_{1\leq r<l}^{N}|z_r-z_l|^{k-2(m_r+m_l+\omega _r \omega _l k/2 +\omega _l m_r+\omega _r m_l) }}{ \prod_{1<r<l}^{M}|w_r-w_l|^{-k}\prod _{t=1}^{N}\prod_{r=1}^{M} |w_r-z_t|^{k -2 m_t}} \times  \nonumber \\ && \times \frac {\prod_{1\leq r<l}^{N}   (\bar{z}_r-\bar{z}_l)^{ m_r + m_l  -\bar{m}_r-\bar{m}_l +\omega _l (m_r- \bar {m}_r) +\omega _r (m_l - \bar {m}_l)}}{\prod_{1<r<l}^{M}     (\bar{w}_r-\bar{z}_t)^{m_t -\bar {m}_t} }   \label{F}
\end{eqnarray}
The parameter $b$ of the Liouville theory is related to the Kac-Moody level $k$ through
\begin{equation}
b^{-2}=k-2 \  \label{cosita},
\end{equation}
while the quantum numbers of the states of both conformal models are also
related each other in a simple way; namely
\begin{equation}
\alpha _i = bj _i + b + b^{-1}/2  \ , \ \ i=\{ 1, 2, ... N \} ,
\end{equation}
The factor $c_k$ in (\ref{N}) is a $k$-dependent ($j$-independent) normalization; see \cite{R}. Furthermore, the following constraints also hold
\begin{eqnarray}
&&\sum _{i=1}^N m _i = \sum _{i=1}^N \bar {m} _i =\frac k2 (N-M-2) \\
&&\sum _{i=1}^N \omega _i = M+2-N \\
&&s = -b^{-1} \sum_{i=1}^N \alpha _i +b^{-2} \frac M2+1+b^{-2} \ . \label{ese}
\end{eqnarray}
In (\ref{ese}), the integer number $s$ refers to the amount of screening operators $V_b (z)= \mu e^{\sqrt {2}b\varphi (z)}$ to be included in Liouville correlators in order to get a non vanishing result. The whole amount of vertex operators involved in the r.h.s. of (\ref{rt}), as mentioned, is
given in terms of the winding numbers of the $AdS_3$ strings; namely
\begin{equation}
\sum _{i=1} ^N \omega _i = M+2-N
\end{equation}
This allows for the possibility of describing scattering amplitudes that, in
principle, violate the winding number conservation.

Notice that, according to (\ref{cosita}), the central charge of the Liouville theory in the r.h.s. of (\ref{rt}) is
given by $c= 1+6Q^2$ with $Q=b+b^{-1}$, while the central charge of the level
$k$ WZNW model is $c=3+6b^{-2}= 3+\frac {6}{k-2}$. Then, if one is interested
in interpreting the relation (\ref{rt}) as stating the equivalence between a
pair of conformal models (and we do have such intention), both central charges should be made to coincide
(presumably by considering an additional internal CFT supplementing the
Liouville action) and, besides, the overall factor $F_k (z_1, ...z_N;w_1, ...w_M)$ should be
suitable to be interpreted as arising from the Wick contractions in the OPE of
such internal theory in the r.h.s. Indeed, the aim of this note is that of
demonstrating that Ribault formula (\ref{rt}) can be actually thought of as a realization of the worldsheet conformal theory
of $AdS_3$ strings in terms of free fields. To be more precise: In the next section we explicitly construct a free field
representation of the $sl(2)_k$ WZNW model in terms of a product of CFTs with
the form
\begin{equation}
Liouville \times U(1) \times time \label{barrita}
\end{equation}
where the factor $time$ refers to a time-like free boson while the $U(1)$
corresponds to a field with non-trivial background charge (see below for
details). Within this framework, the Ribault formula is better understood and,
since it is rigorously deduced from the correspondence between KZ and BPZ
differential equations \cite{S}, enables us to prove the
equivalence between the worldsheet CFT of $AdS_3$ strings and a (tachyonic)
linear dilaton background. The realization we propose here turns out to be reminiscent of the quoted FZZ
conjectured duality and, besides, is closely related to the
discrete light-cone Liouville approach \cite{HHS} as we will demonstrate in
section 3. We will also show the connection with the Wakimoto free field
representation in detail and discuss the computation of WZNW correlation functions
within this framework. Such correlators actually correspond to the scattering amplitudes
of winding strings in the $AdS_3$ space.

In order to avoid redundant introductions, we will employ the standard
notation for describing both WZNW and Liouville conformal models and refer to
the bibliography for details of the nomenclature and conventions.

\subsection{The conformal field theory}

Here, we attempt to construct a CFT theory with the property of realizing the right hand side
of the relation (\ref{rt}). This CFT will take the form (\ref{barrita}),
with the Liouville CFT as a particular factor. The first step for constructing
such realization is supplementing the Liouville action with a pair of
additional bosonic fields $X^0(z)$ and $X^1(z)$ with time-like and space-like
signature respectively. These fields have correlators
\begin{eqnarray}
\langle  \varphi (z_a)  \varphi (z_b) \rangle = \langle  X^1(z_a)  X^1(z_b) \rangle = -\langle  X^0(z_a)  X^0(z_b) \rangle =- 2\log |z_a - z_b| \nonumber
\end{eqnarray}
Moreover, we will demand that field $X^1(z)$ couples to the worldsheet
curvature generating a linear dilaton term in the $U(1)$ direction. In terms of the Liouville field
and its new partners the ``free part'' of the action reads
\begin{equation}
S= \frac {1}{4\pi} \int d^2z \left( - \partial \varphi \bar \partial \varphi +QR\varphi + \mu e^{\sqrt {2}b\varphi }+ \partial X^0 \bar \partial X^0-  \partial X^1 \bar \partial X^1 +i \alpha _0 RX^1  \right) \label{action}
\end{equation}
where the first three terms correspond to the Liouville action, with $Q=b+b^{-
1}$, and the last term represents a second ``background charge'' with $\alpha _0 =
-\sqrt {k}$. On the other hand, a marginal deformation has to be also included in order to
fully represent the r.h.s. of (\ref{rt}). This is due to the presence of the
$M$ additional integrated vertex operators $V_ {-\frac {1}{2b}}(w)$ which, in
our description, turn out to correspond to ``screening
operators''. Besides, the arising of powers of $(w_r -z)$ in (\ref{F}) has to be realized as well. Indeed, if we consider a deformation of action (\ref{action}) given by the
introduction of the marginal term
\begin{equation}
\int d^2z \ \Phi _{aux}(z) \ , \ \ \Phi _{aux}(z) = V_{-\frac {1}{2b}} (z) \times e^{i\sqrt {\frac {1}{2b^2}+1} X^1(z)} \times h.c.= e^{-\sqrt {\frac {k-2}{2}}\varphi (z)+i\sqrt {\frac k2} X^1(z)} \times h.c.  \label{aux}
\end{equation}
then both the factor $F_k(z_1, ...z_N;w_1, ...w_M)$ and the $M$ insertions of operators 
\[
\int d^2w V_{-\frac {1}{2b}}(w) \ , \ \ V_{\alpha}(w) = e^{\sqrt {2}\alpha \varphi (w)}\times h.c. \ ,
\]
naturally arise in (\ref{rt}). It is
easy to verify that $\Phi _{aux}(z)$ is a $(1,1)$-operator with respect to the CFT
described by action (\ref{action}) and stands in the correlators ({\it i.e.} a bunch of them) to screen the background charge $\alpha _0$ (and to help operators $e^{\sqrt {2}\varphi (z)}$ in the task of screening the charge $Q$). The factor $\frac {1}
{M!}$ in (\ref{rt}) is also consistent with the Coulomb gas-like realization
of the correlators due to the multiplicity of permutation of the $M$ screening to be inserted. Furthermore, the conservation laws
\begin{eqnarray}
&&\sum _{i=1}^N \alpha _i -\frac {1}{2b} M+bs = Q  \label{t13} \\
&&\sum _{i=1}^N(m_i+\frac k2 \omega _i) = \sum _{i=1}^N(\bar {m}_i+\frac k2 \omega _i) = 0 \\
&&\sum _{i=1}^N \omega _i = M+2-N\ , \label{conservation}
\end{eqnarray}
that are necessary for the correlators to be non-vanishing, agree with those conditions coming
from the integration over the zero-mode of fields $\varphi (z)$, $X^0(z)$ and
$X^1(z)$. In terms of the quantum numbers $j_i$ and $k$ Eq. (\ref{t13}) reads 
\begin{equation}
\sum _{i=1}^N j_i + N+s-1+\frac {k-2}{2} (N-M-2)=0 \ .  \label{antes}
\end{equation} 
From (\ref{conservation}) we conclude that, in order to conserve the total winding number in $AdS_3$ space, we have to consider the particular case $M=N-2$ and, consequently, 
\begin{eqnarray}
\sum _{i=1}^N j _i +N+s =1 \ .
\end{eqnarray}
This leads to the configuration studied in \cite{RT}. Operator $\Phi _{aux} (z)$ arises in the action of the CFT as being coupled by a constant $\mu / c_k$, according to the KPZ scaling manifested in Eq. (\ref{rt}). Besides, notice that in our convention $c_k$ depends on $\mu $ as well, since the coupling constant of the interaction term $\int d^2z \Phi _{aux}(z)$ should be proportional to $\mu ^{-1/2b^2} = \mu ^{1-k/2}$. Thus, $c_k$ controls the scale where the Liouville potential $e^{\sqrt {2}b\varphi (z)}$ competes with the screening term $e^{-\sqrt {\frac {2}{k-2}}\varphi (z)+i\sqrt {\frac k2}X(z)}$.

By using the coordinates $\rho (z) , \ X^0(z)$ and $X^1(z)$, the vertex operators representing the winding string states in $AdS_3$ would be given by the following expression (cf. (\ref{abo}) below)
\begin{equation}
\Phi_{j,m,\bar m}^{\omega }(z,\bar z) =  \frac {c_k \Gamma (-m-j)}{\Gamma (j+1+\bar m )} V_{\frac {j+k/2}{\sqrt {k-2}}}(z,\bar z) \times e^{i\sqrt {\frac 2k} (m-\frac k2)X^1(z)-i\sqrt {\frac 2k}(m+\frac k2 \omega )X^0(z)} \times h.c.  \label{vertex}
\end{equation}
where $h.c.$ stands for the anti-holomorphic contribution. The normalization $\frac {\Gamma (-m-j)}{\Gamma (1+j+\bar m )}$ is
required to generate the $m$-dependent overall factor $N_k (j_1, ...j_N; m_1, ...m_N)$. Besides, notice that operators (\ref{vertex}) have conformal dimension
\begin{equation}
h= \alpha (Q-\alpha )+(m-\frac k2) +\frac 1k (m-\frac k2)^2-\frac 1k (m+\frac k2 \omega )^2 = -\frac {j(j+1)}{k-2}-m\omega -\frac k4 \omega ^2
\end{equation}
which precisely coincides with the mass-shell conditions for string states in
$AdS_3$, see \cite{MO1}. Then, the full theory turns out to be realized by the
action
\begin{eqnarray}
S = \frac {1}{4\pi} \int d^2z && \left( \partial X^0 \bar \partial X^0-  \partial X^1 \bar \partial X^1 - \partial \varphi \bar \partial \varphi +\frac {k-1}{\sqrt {2(k-2)}} R \varphi -i \sqrt {\frac k2} R X^1 + \right. \nonumber \\ 
&& \left. + \frac {\mu }{ \ c_k } e^{-\sqrt {\frac {k-2}{2}}\varphi +i\sqrt {\frac k2} X^1}+ \mu e^{\sqrt {\frac {2}{k-2}}\varphi } \right) \label{S}
\end{eqnarray}
and corresponds to the following stress-tensor
\begin{equation}
T(z) =\frac 12 (\partial X^0 (z))^2 -\frac 12 (\partial X^1 (z))^2 -i\sqrt{\frac k2} \partial^2 X^1(z) -\frac 12 (\partial\varphi (z))^2 + \frac {k-1}{\sqrt{2(k-2)}} \partial ^2 \varphi (z)  \label{T}
\end{equation}
Hence, including the contributions of both fields $X^0(z)
$ and $X^1(z)$, the central charge now results
\begin{equation}
c= c_{\varphi}+c_{X^0,X^1} = 1+6Q^2+2-6k= 3+\frac {6}{k-2}, \label{c}
\end{equation}
and this exactly coincides with the central charge of the $SL(2,R)_k$ WZNW model. The free field theory actually presents a $\hat {sl(2)}_ k \times \hat {sl(2)}_ k$ symmetry in the sense that the stress tensor $T(z)$ can be constructed by means of the Sugawara form starting with a set of $\hat {sl(2)} _k$ currents (see below). However, it is worth mentioning that such symmetry of the whole theory is not obvious due to the presence of the Liouville cosmological term and the particular representation employed for describing the vertex operators. We can represent that algebra by means of the fields introduces above; namely
\begin{eqnarray}         
J^{+} (z) &=& \left( -i \sqrt {\frac k2} \partial X^1 (z) + \sqrt {\frac {k-2}{2}}\partial \varphi (z)  \right) e^{i\sqrt {\frac 2k} (X^0(z)-(k-1) X^1(z))+\sqrt {2(k-2)}\varphi (z)} \nonumber \\
J^{-} (z) &=& \left( -i \sqrt {\frac k2} (2k-3) \partial X^1 (z) + \sqrt {\frac {k-2}{2}}(2k-1)  \partial \varphi (z)  \right) e^{-i\sqrt {\frac 2k} (X^0(z)-(k-1) X^1(z))+\sqrt {2(k-2)}\varphi (z)}  \nonumber \\
J^3(z) &=& -i\sqrt {\frac k2}\partial X^0 (z) \label{corrientes}
\end{eqnarray}
and their anti-holomorphic analogues. These currents realize the appropriate OPE; namely
\begin{eqnarray}
J^3(z)J^{\pm }(w) &=& \pm \frac {1}{(z-w)} J^{\pm } (w) +... \\
J^3(z)J^{3}(w) &=& - \frac {k/2}{(z-w)^2}  +... \\
J^-(z)J^{+}(w) &=& \frac {k}{(z-w)^2}  - \frac {2}{(z-w)} J^{3 } (w) +...
\end{eqnarray}
where the dots ``$ ...$'' represent regular terms in the short distance expansion. This means that the Fourier modes $J_n^a$, defined as
\begin{equation}
J^a _n = \frac {1}{2\pi i} \oint dz \ z^{-1-n} J^a(z)
\end{equation}
with $a= \{ 3,+,-\} $, satisfy the $sl(2)_k$ affine Kac-Moody algebra, namely
\begin{eqnarray}
&& [J^-_n,J^+_m]= -2 J^{3}_{n+m} + nk\delta _ {n,-m} \nonumber \\
&& [J^3_n,J^{\pm }_m]= \pm  J^{\pm }_{n+m} \nonumber \\
&& [J^3_n,J^3_m]= - n\frac k2 \delta _ {n,-m} \nonumber
\end{eqnarray}
Then, our free field realization is now complete, and it enables to understand
the Ribault formula (\ref{rt}) in a natural way; namely: as a simple realization of $AdS_3
$ string theory in terms of a linear dilaton background with tachyon-type interactions represented by ($1,1$)-exponential fields (\ref{aux}). Moreover, due to the
closed relation between the non linear $\sigma $-model in both $AdS_3$ and in
2D black hole, this realization turns out to be reminiscent of the FZZ
duality. In the next section we discuss this feature as well as the relation
with other approaches based on linear dilaton description. It is worth mentioning that other interesting relations between two dimensional CFts and the theory with $\hat {sl(2)}_k$ symmetry can be found in the literature. One of these, which in our opinion deserves particular attention, is the one presented in Ref. \cite{PN}, where it is shown how to add appropriate ghost fields to an arbitrary CFT and we use them to construct the $\hat {sl(2)}_k$ currents. It would be interesting to understand whether there exists a relation between our free field realization and such kind of embedding.

\subsection{Free field representation}

Here, as mentioned, we address the problem of establishing a connection between the
free field realization presented above and those formalisms that, following
analogous lines, were employed in the literature to represent string theory in
$AdS_3$. For instance, a natural question arises as to whether is a clear
connection between the realization we presented here and the quoted Wakimoto
free field representation of the $\hat {sl(2)}_k$ current algebra. But, before
analyzing such a relation, we find convenient to comment on a second approach,
called the discrete light-cone Liouville theory.

In Ref. \cite{HHS}, Hikida, Hosomichi and Sugawara ``demonstrate[d] that string
theory [on $AdS_3$ background] can be reformulated as a string theory defined
on a linear dilaton background''. Our aim here is to show that the ``approach of
free field realization'' of \cite{HHS} is related to ours in a very simple way
which, basically, corresponds to perform a $U(1)$ transformation on the fields
and replace the marginal deformation in the action. Namely, let us define new coordinates by means of the rotation
\begin{eqnarray}
&& \rho (z) = (1-k) \varphi (z) + i\sqrt {k(k-2)} X^1(z) \label{acordate1} \\
&& Y^1(z) =  i\sqrt {k(k-2)} \varphi (z)+ (k-1) X^1 (z) \label{newc}
\end{eqnarray}
and simply rename 
\begin{equation}
Y^0(z)= -X^0(z),
\end{equation} 
This rotates the background charge as
\begin{equation}
\frac {Q}{\sqrt {2}} \varphi (z)-i\sqrt {\frac k2} X^1(z) = -\frac {1}{\sqrt {2(k-2)}} \rho (z) \ ,
\end{equation}
Hence, the dilaton field only depends on the new coordinate $\rho (z)$ and the theory becomes strongly coupled in the large $\rho (z) $ region. In these coordinates, the currents (\ref{corrientes}) take the usual form that appears in the standard bosonization of parafermions; namely
\begin{eqnarray}         
&& J^{\pm } (z) = \left( - i \sqrt {\frac k2} \partial Y^1 (z) \pm \sqrt {\frac {k-2}{2}}\partial \rho (z)  \right) e^{\mp i\sqrt {\frac 2k} (Y^0(z)+Y^1(z))} \\
&& J^3(z) = i\sqrt {\frac k2}\partial Y^0 (z) \label{corrientess}
\end{eqnarray}
Then we find that the Sugawara stress tensor takes the form
\begin{equation}
T(z) = -\frac 12 (\partial\rho (z))^2 - \frac {1}{\sqrt{2(k-2)}} \partial ^2 \rho (z) -\frac 12 (\partial Y^1 (z))^2 + \frac 12 (\partial Y^0 (z))^2 \ ,  \label{TT}
\end{equation}
which precisely corresponds to the free field realization discussed in Ref. \cite{HHS}. In that paper, the interaction term to be added to the CFT (\ref{TT}) in order to fully describe the non-linear $\sigma $-model on $AdS_3$ was given by the ($1,1$)-operator
\begin{equation}
\partial Y^1_L(z) \bar {\partial } Y^1_R(\bar z) e^{-\sqrt {\frac {2}{k-2}}\rho (z,\bar z)}   \label{graviton}
\end{equation}
instead of $\Phi _{aux} (z)$ in (\ref{aux}). It is important mentioning that the action containing the interaction term (\ref{graviton}) and its duality to the $N=2$ Liouville (and sine-Liouville eventually) was also observed in Ref. \cite{coreanos} and further clarified in Ref. \cite{Yu}, where its connection to the $SL(2,R)_k/U(1)$ WZNW model is discussed. This was also studied in \cite{Satoh2} as a particular screening charge of the model.

Furthermore, the new coordinates (\ref{acordate1})-(\ref{newc}) are also useful to discuss another realization, based on the FZZ duality. In fact, in terms of the fields $\rho (z) , \ Y^1(z)$ and $Y^0 (z)$, the sine-Liouville theory is obtained by replacing the graviton-like marginal deformation (\ref{graviton})
by the operator
\begin{equation}
e^{-\sqrt {\frac {k-2}{2}}\rho (z)} \left( e^{i\sqrt {\frac k2}(Y_L ^1(z)-Y_R ^1(\bar z ))}+e^{i\sqrt {\frac k2}(Y^1_R (\bar z )-Y^1_L (z))}\right) = 2 e^{-\sqrt {\frac {k-2}{2}}\rho (z)}\cos \left( \sqrt {k/2} (Y^1_L (z)-Y^1_R (\bar z ))\right) , \label{taquion}
\end{equation}
where $Y_L ^1(z)$ and $Y_R ^1(\bar z )$ refer to the left and right modes of the field respectively (cf. (\ref{psssi}) below). It is worth emphasizing that (\ref{graviton}) and (\ref{taquion}) represent two
different marginal deformations of the same linear dilaton backgrounds which,
besides, preserve the $\hat {sl(2)}_k \times \hat {sl(2)}_k $ symmetry, as it
can be verified from the OPE with currents (\ref{corrientes}).

At this point, we are able to show the explicit connection with the Wakimoto free field representation as well. It follows from a (de)bosonization procedure. In fact, the Wakimoto free fields appear once we define
\begin{eqnarray}
\beta (z) &=& J^-(z) \ , \ \ \gamma (z) =  e^{-i\sqrt {\frac 2k} (Y^0 (z)+Y^1(z))}  \\
\phi (z) &=& \rho (z) +i\sqrt {\frac {k-2}{k}} (Y^0(z)+Y^1(z))
\end{eqnarray}
This leads to find the standard representation for the stress-tensor, namely
\begin{equation}
T (z) = -\frac 12 (\partial \phi (z) )^2 -\frac {1}{\sqrt {2(k-2)}} \partial ^2 \phi (z) -\beta (z) \partial \gamma (z)
\end{equation}
which, of course, corresponds to the Sugawara construction by employing (\ref{corrientes}). Regarding the interaction term (written in this language) the authors of \cite{HHS} showed that (\ref{graviton}) agrees with the
usual form
\[
           \beta (z) \bar {\beta } (\bar z ) e^{-\sqrt{\frac {2}{k-2}}\phi (z)}
\]
for the screening operator of the non-linear $\sigma $-model on $AdS_3$ \cite{B,GN3}, up to a total derivative of the zero-dimension operator $e^{-\sqrt {\frac {2}{k-2}}\rho (z)}$.

It is also worth pointing out that the relation between both marginal deformations (\ref{graviton}) and (\ref{taquion}) is similar to the one existing between the ``standard'' and the ``conjugate'' representations of the $SL(2,R)_k/U(1) \times U(1)$ WZNW model discussed in Ref. \cite{GN3,GL,GN2}. While (\ref{taquion}) is associated to the representations $\Phi ^{\omega }_{j,m,\bar {m}} (z)$ employed in \cite{BB}, the ``screening'' of the form (\ref{graviton}) corresponds to the conjugate representations $\tilde {\Phi }^{\omega }_{j,m,\bar m } (z)$ which are associated to the discrete states of the 2D black hole \cite{GL}.

\section{String theory on $AdS_3 \times {\cal N}$}

\subsection{The spectrum of the free theory}

The string theory on Euclidean $AdS_3$ is described by the $SL(2,C)_k/SU(2)$ WZNW model, while the Lorentzian version is constructed in terms of its analytic extension to the $SL(2,R)_k$ WZNW, which is certainly less understood in what respects to the formal aspects. Within the context of string theory, the Kac-Moody level $k$ of the WZNW theory turns out to correspond to the quotient between the typical string length scale $l_s$ and the ``$AdS$ radius'' $l \sim (-\Lambda ) ^{-1/2}$, namely $k \sim l^2 / l_s^2 $. Consequently, $k$ controls the coupling of the theory leading, in the large $k$ limit, to both classical and flat limit.

The Hilbert space of this theory is then given in terms of certain representations of $SL(2,R)_k \times SL(2,R)_k$. Then, the string states are described by vectors $\left| \Phi ^{\omega }_{j,m,\bar m}\right>$ which are defined by acting with the vertex operators $\Phi ^{\omega }_{j,m,\bar m}(z)$ on the $SL(2,R)_k$-invariant vacuum $\left| 0 \right>$; namely
\[
\lim _{z\to 0 } \Phi^{\omega }_{j,m,\bar m } (z) \left| 0 \right> = \left| \Phi ^{\omega }_{j,m,\bar m}\right>
\] 

In order to define the string theory precisely, it is necessary to identify which is the subset of representations that have to be taken into account. Such a subset has to satisfy several requirements: In the case of the free theory these requirements are associated to the normalizability and unitarity of the string states. At the level of the interacting theory, additional properties are requested, as the closeness of the fusion rules, the factorization properties of $N$-point functions, etc.

Even in the case of the free string theory, the fact of considering non-compact curved backgrounds is not trivial at all. The main obstacle in constructing the space of states is the fact that, unlike what happens in flat space, in curved space the Virasoro constraints are not enough to decouple the negative-norm string states. Then, in the early attempts for constructing a consistent string theory in $AdS_3$, additional {\it ad hoc} constraints were imposed on the vectors of the $SL(2,R)_k$ representations. Usually, the vectors of $SL(2,R)$ representations are labeled by a pair of indices $j$ and $m$, and such additional constraints, demanded as sufficient conditions for unitarity, implied an upper bound for the index $j$ of certain representations (namely, the discrete ones and consequently for the mass spectrum). The modern approaches to the ``negative norm states problem'' include such a kind of constraint as well, although it does not implies a bound on the mass spectrum as the old versions do; see \cite{MO3, GKS, KS, Oo, EGP} for details. The mentioned upper bound for the index $j$ of discrete representations, often called ``unitarity bound'', reads 
\begin{equation}
\frac {1-k}{2}<j<-\frac 12.  \label{unitarity}
\end{equation}
In the case of Euclidean $AdS_3$, the spectrum of string theory is given by the continuous series of $SL(2,C)$, parameterized by the values $j=-\frac 12 +i\lambda $ with $\lambda \in R$ and by real $m$. Besides, the case of string theory in Lorentzian $AdS_3$ is richer and its spectrum is composed by states belonging to both continuous ${\cal C}_{\lambda }^{\alpha ,\omega }$ and discrete ${\cal D}^{\omega ,\pm}_{j}$ series. The continuous series  ${\cal C}_{\lambda }^{\alpha ,\omega }$ have states with $j=-\frac 12 +i\lambda $ with $\lambda \in R$ and $m-\alpha \in Z$, with $\alpha \in [0,1) \in R$ (as in $SL(2,C)$, obviously). On the other hand, the states of discrete representations ${\cal D^{\pm, \omega}}_{j}$ satisfy $j = \pm m-n$ with $n\in Z_{\geq 0}$. So, the next step is explaining the index $\omega $: In order to fully parameterize the spectrum in $AdS_3$ we have to introduce the ``flowed'' operators $\tilde {J}^{a}_n$ ($a\in \{ 3,-,+\} $), which are defined through the spectral flow automorphism
\begin{eqnarray}
&& J^3 _n\to \tilde {J}^3 _n = J^3 _n + \frac k2 \omega \delta _{n,0} \\
&& J^{\pm } _n\to \tilde {J}^{\pm} _n = J^{\pm } _{n\pm \omega } \label{arrova}
\end{eqnarray}
acting of the original $\hat{sl(2)}_k$ generators $J ^a_n$. Then, states $\left| \Phi _{j,m,\bar m}^{\omega }\right>$ belonging to the discrete representations ${\cal D}^{\pm , \omega }_j$ are those obeying
\begin{eqnarray}
\tilde {J}^{\pm }_0 \left| \Phi _{j,m,\bar m}^{\omega }\right> &=& (\pm j-m) \left| \Phi _{j,m\pm 1,\bar m}^{\omega }\right> \label{t1} \\
\tilde {J}^{3}_0 \left| \Phi _{j,m,\bar m}^{\omega }\right> &=& m \left| \Phi _{j,m, \bar m}^{\omega }\right> \label{t2}
\end{eqnarray}
and being annihilated by the positive modes, namely
\begin{eqnarray}
\tilde {J}^a_n  \left| \Phi _{j,m,\bar m}^{\omega }\right> =0 \ , \ \ \ n>0 \ ; \label{t3}
\end{eqnarray}
and analogously for the antiholomorphic modes. Notice that states with $m=\pm j$ represent highest (resp. lowest) weight states. On the other hand, primary states of the continuous representations ${\cal C}^{\alpha , \omega }_{\lambda }$ turn out to be annihilated by all the positive modes, as in (\ref{t3}), though for any zero-mode $J^a_0$. The excited states in the spectrum are defined by acting with the negative modes $J^a_{-n}$ ($n\in Z_{>0}$) on the Kac-Moody primaries $\left| \Phi _{j,m,\bar m}^{\omega }\right>$; these negative modes play the role of creation operators ({\it i.e.} creating the string excitation). The ``flowed states'' (namely those being primary vectors with respect to the $\tilde {J}^a_n$ defined with $|\omega |>1$) are not primary with respect to the $\hat {sl(2)}_k$ algebra generated by $J^a_n$, and this is clearly because of (\ref{arrova}). However, highest weight states in the series ${\cal D}^{+,\omega }_j$ are identified with lowest weight states of ${\cal D}^{-,\omega }_{-k/2-j}$, which means that spectral flow with $|\omega |=1$ is closed among certain subset of Kac-Moody primaries.

The states belonging to discrete representations have a discrete energy spectrum and represent the quantum version of those string states that are confined in the centre of $AdS_3$ space; these are called ``short strings'' and are closely related to the states that are confined close to the tip in the 2D Euclidean black hole geometry. On the other hand, the states of the continuous representations describe massive ``long strings'' that can escape to the infinity because of the coupling with the NS-NS field. For the long strings, the quantum number $\omega $ can be actually thought of as a winding number asymptotically. 

The vertex operators of the theory on $AdS_3\times {\cal N}$ are then given in terms of the $SL(2,R)_k$ representations and take the form
\begin{equation}
\Phi _{j,m,\bar m, q}^{\omega } (z) = \Phi _{j,m,\bar m}^{\omega } (z) \times V_q (z)
\end{equation}
where $V_q (z)$ represents the vertex operator on the ``internal'' CFT defined on ${\cal N}$. The Virasoro constraint $L_0 = 1$ then implies that the conformal dimension of the fields satisfies
\begin{equation}
h_{(j,m,q)} = -\frac{j(j+1)}{k-2}-m\omega -\frac k4 \omega ^2+h_{{\cal N}}(q) =1
\end{equation}
where $h_{{\cal N}} (q)$ is the conformal dimension of $V_q (z)$. This agrees with the functional form we obtained in section 2. Regarding the value of $k$, this is given by $k=2(26-c_{{\cal N}})/(23-c_{{\cal N}})$, where $c_{{\cal N}}$ is the central charge on ${\cal N}$.

Besides those we commented above, there are more unitary representations which, in principle, could be also included in the definition of the Hilbert space: For instance, we have the complementary series ${\cal E}_{\alpha }^{j}$ with $-1 <j <-\frac 12$, $-\frac 12 -j <|\alpha -\frac 12|$ and $m-\alpha \in Z$. It is usually assumed that these representations are not explicitly (but implicitly) present in the space of states since the continuous and discrete series form a complete basis of the square integrable functions on $AdS_3$ (and thus all additional states would be already taken into account as certain linear combinations of states in ${\cal D}^{\pm , \omega }_j$ and ${\cal C}^{\alpha , \omega }_ {\lambda }$). However, this argument is valid in the large $k$ regime, where the semi-classical arguments based on the differential functions on the space make sense. For instance, in order to conclude that the complementary series are to be excluded it would be necessary to show that, in addition, such states do not decouple when taking the large $k$ limit in the operator product expansion: If those state do not decouple then it would be necessary to exclude them in the definition of the Hilbert space. Conversely, if their do decouple in the semi-classical limit, then the argumentation in terms of the composition of the Hilbert space in the large $k$ limit is not enough for excluding them {\it ab initio} and they could be actually necessary. However, there exists more evidence in favour of the idea that the whole spectrum of the theory is constructed just by the continuous and discrete series; this is suggested by the modular invariance of the one-loop partition function in thermal $AdS_3$ (studied in Ref. \cite{MO2}) and the fact that only long strings and short strings do appear in the spectrum.

Besides, other states, having momentum $j = -((n+1)(k-2)+1)/2$ with $n\in Z_{\geq 0}$, also appear in the theory. These are certainly non-perturbative states since, for heavy states, they have masses of the order $j$ and, hence, of the order of the string tension. Since $k$ controls the semi-classical limit, these states would decouple from the perturbative spectrum when $k$ goes to infinity. These states are associate to worldsheet instantons and the semi-classical interpretation of them relates their presence with the emergence of non-local effects in the dual CFT on the boundary (according to the $AdS/CFT$ correspondence).

\subsection{Discrete light-cone Liouville theory}

Now, we discuss the indices $j$, $m$ and $\omega $ of $SL(2,R)_k$ representation in the context of the free field description.

In our realization we were considering operators of the form
\begin{equation}
\Phi ^{\omega }_{j,m,\bar m}(z) \propto e^{\sqrt {\frac {2}{k-2}} (j+\frac k2)\varphi (z) +i\sqrt {\frac 2k}(m-\frac k2)X^1(z)-i\sqrt{\frac 2k }(m+\frac k2 \omega )X^0(z)} \ , \label{abo}
\end{equation}
which can be actually thought of as operators on the product $\frac {SL(2,R)_k}{U(1)}\times time$, where the time-like U(1) is parameterized by the bosonic field $X^0(z)$. The part of the vertex describing the theory on the coset has conformal dimension
\begin{equation}
h_{(j,m)}=-\frac {j(j+1)}{k-2}+\frac {m^2}{k}
\end{equation}
Then, we notice that this formula remains invariant under the involution
\begin{eqnarray}
&& j \to -j(k-1)-m(k-2)-k/2  \label {invo} \\
&& m \to jk+m(k-1)+k/2  \label{invo2}
\end{eqnarray}
In order to extend this symmetry to the whole theory ({\it i.e.} to the theory on the product $\frac {SL(2,R)_k}{U(1)}\times U(1)$) we have to consider a transformation on the spectral flow parameter $\omega $ as well; namely
\begin{equation}
\omega \to -\omega -1-2(j+m) \ . \label{invo3}
\end{equation}
Performing transformation (\ref{invo})-(\ref{invo3}) on (\ref{abo}), we get
\begin{equation}
V_{j,m,\bar m, \omega}(z) = e^{-\sqrt {\frac {2}{k-2}} (j(k-1)+m(k-2))\varphi (z) +i\sqrt {\frac 2k}(jk+m(k-1))X^1(z)-i\sqrt{\frac 2k}(m+\frac k2 \omega )X^0(z)} \label{abo2}
\end{equation}
and, expressed in terms of fields $\rho (z)$, $Y^0(z)$ and $Y^1(z)$, these operators become
\begin{equation}
V_{j,m,\bar m ,\omega} = e^{\sqrt {\frac {2}{k-2}}j\rho (z)-i\sqrt {\frac 2k}mY^1(z)-i\sqrt {\frac 2k}(m+\frac k2 \omega )Y ^0(z)}   \ ,
\end{equation}
which are precisely those that were considered in the discrete light-cone Liouville approach to string theory in $AdS_3$ discussed in \cite{HHS}. In Ref. \cite{HHS} the authors proposed that the winding number of strings in $AdS_3$ can be seen as a quantized momentum in the direction $(Y^1-Y^0)/ \sqrt {2}$ which, according to the discrete light-cone prescription, turns out to be compactified with period $2\pi \sqrt {2/k}$; namely
\begin{equation}
Y^0 (z)- Y^1 (z) =  Y^0 (z)- Y^1 (z) + \frac {4\pi n}{\sqrt{k}} , \ \ n \in Z \ .
\end{equation}
Within this framework, the index $\omega $ corresponds to the topological winding along the compact light-like direction, and in terms of the Liouville field $\varphi (z)$ this direction reads
\begin{equation}
X^0 (z)- X^1 (z) +  k \left(  i\sqrt {\frac {k-2}{k}}\varphi (z)-X^1(z)    \right) \ .
\end{equation} 
Now, the question is: what do the quantum numbers $j$ and $m$ mean in terms of this free field representation? To understand this, let us comment on the $SL(2,R)_k$ representations again: As already mentioned, the principal continuous series ${\cal C}^{\alpha , \omega }_{\lambda }$ correspond to $j=-\frac 12 +i\lambda $ with $\lambda \in R$ and, through (\ref{invo}), this results in the new values $j= -\frac 12 +i\tilde {\lambda }-m(k-2)$ with $\tilde {\lambda }\in R$, which only seem to belong to the continuous series if $m=0$. Besides, (\ref{invo2}) implies that, after performing the change (\ref{invo2}), $m$ turns out to be a non-real number. Then, the relation between (\ref{abo}) and (\ref{abo2}) can not be thought of as a simple identification between states of different continuous representations but it does correspond to different free field realizations (at least in what respects to the continuous series ${\cal C}^{\alpha , \omega}_{\lambda }$). On the other hand, regarding the discrete representations, it is worth mentioning that the quantity $j+m$ remains invariant under the involution (\ref{invo})-(\ref{invo2}); though it is not the case for the difference $j-m$. Unlike the states of continuous representations, transformation (\ref{invo})-(\ref{invo3}) is closed among certain subset of states of discrete representations. This is because such transformation maps states of the discrete series with $2(j+m)\in Z$ among themselves. In particular, the case $m+j=0$ corresponds to the well known identification between discrete series ${\cal D}^{\pm , \omega =0}_{j}$ and ${\cal D}^{\mp , \omega =\pm 1}_{-k/2-j}$ since in that case (\ref{invo})-(\ref{invo3}) reduce to $j\to -k/2-j, \ m\to k/2 -m=k/2-j , \ \omega \to -1-\omega$. The relation between quantum numbers manifested by (\ref{invo}) and (\ref{invo2}) permits to visualize the relation between the vertex considered in our construction and those of reference \cite{HHS} and consequently relate the free field realization (\ref{S}) with the one of the discrete light-cone approach. The relation between both is a kind of ``twisting'' and is closely related to the representations studied in \cite{GL}. Other difference regards the interaction: Unlike the free field representation based on action (\ref{S}), the one of \cite{HHS} comes from a field redefinition (through a bosonization and a $SO(1,2)$ transformation after of it) of the non-linear $\sigma $-model on $AdS_3$. Thus, the action employed in the discrete light-cone realization of $AdS_3$ strings involves the interaction term (\ref{graviton}). This makes an important difference when trying to compute correlation functions. In \cite{HHS} it was discussed how the interaction term (\ref{graviton}) plays an important role for the emergence of short string states in the spectrum. Within such framework, such states are an effect due to the interaction term. 

In the following subsections we study what happens when the interactions are taken into account.

\subsection{The spectrum of the interacting theory}

When interactions are considered, additional restrictions to those imposed for unitarity on
the free spectrum may appear. This is because, besides the requirement of normalizability and
unitarity, it is necessary to guarantee that the fusion rules result closed among
the unitary spectrum. Otherwise, it could be possible to produce negative
norm states by scattering processes involving unitarity incoming states. This
problem was addressed in \cite{P, GN1, MO3, MO2}. After integrating the $N$-point correlators over the worldsheet insertions
$z_a$ ($N\in \{ 1,2, ... N\} $), the pole structure of scattering amplitudes and
the factorization properties of them could also result in additional
constraints on the external (incoming and outgoing) states, and in some
examples such constraints can be even more restrictive than those required for
unitarity. For instance, in the case of string theory in $AdS_3$ the $N$-point
scattering amplitudes are known to be well defined only if the external momenta
of the states involved in the process satisfy
\begin{equation}
\sum _{a=1}^N j_a >3-N-k  \label{tttt}
\end{equation}
This restriction is stronger than the unitarity bound (\ref{unitarity}) imposed
on each state. In Ref. \cite{MO3}, the condition (\ref{tttt}) was analyzed from
the viewpoint of the AdS/CFT correspondence and it was concluded that only
those correlators satisfying such constraint would have a clear interpretation
in terms of a local CFT on the boundary.

We explicitly compute the interactions in the next subsection, where we pay particular attention to the integral representation of correlation functions and its factorization properties.

\subsection{Four-point function}

Now, we move to the study of the factorization with the attention focussed on the role played by the winding states. Let us begin by considering the integral representation of the four-point function; namely
\begin{eqnarray}
\langle \prod _{a=1} ^4 \Phi _{j_a,m_a,m _a}^{\omega _a = 0}(z_a) \rangle &=& \frac {b}{\pi ^{5}} (\pi ^2\mu b^{-2})^{-s} \prod _ {a=1} ^4 \frac {\Gamma (-m_a-j_a)}{\Gamma (j_a+\bar m _a+1)} \prod  _{i=1}^2 \int d^2w_i \prod  _{r=1}^s \int d^2v_r   | w_1  - w_2  |^{2} \nonumber \\  
&& \prod _{a<b}^{4}   | z_a  - z_b  |^{-2(m_a+m_b)-\frac {4}{k-2}(j_aj_b+\frac k2 (j_a+j_b+1))}  \prod _{a=1}^{4} \prod _{r=1}^{s}  |z_a   - v_r  |^{-2-\frac {2}{k-2}(j_a+1)} \times \nonumber \\
&& \times  \prod _{a=1}^{4} \prod _{i=1}^{2}  | z_a  - w_i  |^{2(m_a+j_a)}    \prod _{r=1}^{s}  \prod _{i=1}^{2}  |v_r   - w_i  |^{2}      \prod _{r<t}^{s} | v_r  - v_t  |^{\frac {4}{2-k}} \label{fggg}
\end{eqnarray}
Actually, this is the Coulomb gas-like representation of the Teschner-Ribault where, for simplicity, we considered the case $m_a = \bar m_a$ ($a= 1,2,3,4$), while the general case with $m-\bar m\in Z$ and generic $N$ is discussed in the following subsection for the case of sine-Liouville theory in more detail. Besides, we will consider the case where $k$ is large enough (though finite) to allow configurations of the form $-2j_a \in N$ within the unitarity bound (\ref{unitarity}). This case allows for the correlators to be ``resonant'' in the sense that can be realized by using an integer numbers of screening charges since we assume here the particular cases $-\sum _{a=1}^4 j_a \in Z_{>0}$. In (\ref{fggg}), the inserting points $w_i$ refer to those where the screenings of the kind $\Phi_{aux}(w)$ are inserted, while the points $v_r$ are the locations of the screenings $V_b(v)$.

Now, let us consider the coincidence limit of a pair of external states (let us say $z_2 \to z_1 $ renaming $z_1=0$ and $z_2= \epsilon $). This enables to study the limit where the four-point amplitude factorizes into a pair of three-point sub-amplitudes, and the mass-shell conditions of the intermediate states that are interchanged between both subprocesses arise as poles developed in the coincidence limit $\epsilon \to 0$. In order to analyze this factorization limit, we should also sum over the different ways of taking the limit and, in principle, take $\tilde s$ operators of the form $V_b(v)$ and $p$ operators of the form $\Phi _ {aux} (w)$ going to $z_1=0$ as well (with all the possible $\tilde s$ and $p$; $0\leq \tilde s \leq s$ and $0\leq p\leq 2$ since those are the total amount of screening operators that are available). To parameterize the limit, it is convenient to rename the inserting points as $v_r = v_r /z_2$ for those with $r\leq \tilde s$ and $w_i=w_i /z_2$ for those with $i\leq p$. By explicitly taking the limit $z_2 \to z_1 = 0$ and collecting the powers of $|z_2|$ we find that the four-point correlator (\ref{fggg}) factorizes into two pieces: a three-point correlator in the left hand side containing $p$ screenings $\Phi _ {aux}(w)$ and $\tilde s$ screenings $V_b(v)$, and a second three-point function in the right hand side  which is composed by $2-p$ screenings $\Phi _ {aux}(w)$ and $s-\tilde s$ screenings of the kind $V_b(v)$. The intermediate states, which are interchanged in the channels of the four-point string scattering process, satisfy the mass-shell conditions that can be read of as pole conditions of the propagator; these pole conditions are extracted from the divergences after integrating over $\int d^2z_2$. Hence, as mentioned, such mass-shell conditions can be simply read from the exact value of the exponent $\eta _p$ of $|z_2|^{2 \eta _p}$; and this power turns out to be
\begin{equation}
\eta _p = (p-1)(m_1+m_2)+p(j_1+j_2)+\delta _ {p,2} +(p-1) \tilde s +p+ \frac {2}{k-2}(j_1j_2+\frac k2 (j_1+j_1+1)+\tilde s (j_1+j_2+2) -\frac 12 \tilde s (\tilde s -1))
\end{equation}
The case $p=1$ corresponds to the factorization of (\ref{fggg}) in terms of a pair of three-point sub-amplitudes conserving (each of them and the total one) the winding number. This is because the $N$-point Liouville amplitudes with $N-2$ additional operators $\Phi _{aux}(w)$ represent WZNW amplitudes conserving $\sum _ {a=1}^N\omega _a$. On the other hand, the case $p=0$ corresponds to the case where the conserving winding correlator (\ref{fggg}) splits into the sum of a pair of violating winding correlators. This is because the three-point function in the left hand side has no screening of the kind $\Phi_{aux}(v)$ (a deficit) while the three-point function of the right hand side has two of them (an excess). 

In the case $p=1$, the factor $|z_2|^{2\eta _p}$ standing in the limit $z_2=\epsilon \to 0$ develops poles located at $\eta _1 = -n$ with $n\in Z_{\geq 0}$. This conditions can be written in the following convenient way
\begin{equation}
h_j = -\frac {j(j+1)}{k-2}+n=1 \ ,\ \ \ j=j_1+j_2+\tilde s +1 =0
\end{equation}
where $\tilde s = \{ 0,1,2, ... s\} $. This precisely agrees with the mass-shell condition of an intermediate state belonging to the $SL(2,R)_k$ representation with quantum numbers 
\[
m=m_1+m_2 \ , \ \ \ \ \omega =0 \ .
\] 
Analogously, the poles arising in the case $p=0$ are located at $\eta _0 = -n$ and this can also be written in a similar way; namely
\begin{equation}
h_j = -\frac {j(j+1)}{k-2}-m -\frac k4 +n=1 \ ,\ \ \ j=j_1+j_2+\tilde s +\frac k2 =0  \label{cocho}
\end{equation}
where $m$ and `$\omega $ are now given by 
\[
m=m_1+m_2-\frac k2 \ , \ \ \ \ \omega =1 \ . 
\]
Again, Eq. (\ref{cocho}) represents the mass-shell condition for the intermediate state belonging to the ``flowed'' $SL(2,R)_k$ representations. The case $p=2$ is similar and shows that the amplitude (\ref{fggg}) factorizes in such a way that the poles arising in the limit where two states of the four coincide can be naturally interpreted as mass-shell conditions of states of ``flowed'' $SL(2,R)_k$ representations with $\omega =0 $ and $\omega = 1 $.

The factorization of four-point function was also studied in Ref. \cite{MO3} by
using the usual $x$-basis (by Fourier transforming in $m$). There, it was
shown
that the internal channels of the amplitudes contain contributions of both
$\omega =0$ and $\omega =1$ winding states for certain processes and this is consistent with our result here (see section 4 of \cite{MO3} for details). Moreover,
it
was claimed in \cite{MO3} that only the four-point functions satisfying the constraint
\begin{equation}
j_1+j_2>-\frac {k+1}{2} \ , \ \ j_3+j_4>-\frac {k+1}{2}  \label{copado}
\end{equation}
present a well behaved factorization limit enabling to interpret the pole
structure as corresponding to physical state contributions in the intermediate
channels. In fact, those correlation functions which do not satisfy (\ref{copado}) receive contributions from poles located at
\begin{equation}
j+j_1+j_2+k=-n \ , \ \ n\in Z_{\leq 0} \ .
\end{equation}
and these are not suitable for natural interpretation. A deeper understanding of the additional conditions for the factorization in $AdS_3$ is a non trivial aspect and requires further study indeed. 

Now, we will compare the features of the free field representation discussed above with their analogues for the case of sine-Liouville field theory.

\subsection{Relation with sine-Liouville field theory}

Now, let us return to the tachyonic-type interaction term (\ref{taquion}). One of the most interesting aspect of the relation between $AdS_3$ string
theory and the CFT on linear dilaton background (\ref{S}) is the fact that it looks very
much like the FZZ duality, which associates the 2D black hole ({\it i.e.} the
$SL(2,R)_k/U(1)$ WZNW model) to the sine-Liouville field theory. Indeed, the
realization (\ref{T}) turns out to be reminiscent of such duality due to the presence
of the exponential (tachyonic) term $\Phi _{aux} (z)$, being
\begin{equation}
\Phi _{aux} (z) = e^{-\sqrt{\frac {k-2}{2}}\rho (z)+ i\sqrt {\frac k2}Y^1(z)} \times h.c. \ .  \label{uno}
\end{equation}
Operator $\Phi _{aux} (z)$ is actually ``a half'' of the sine-Liouville interaction term (see (\ref{taquion}) and (\ref{psssi}) below and notice that it has the appropriate exponent). On the other hand, instead of $\Phi _{aux}(z)$, we could indistinctly include in (\ref{S}) a slightly different interaction term; namely: its complex conjugate $\Phi ^* _{aux}(z)$. Since the action is quadratic in the field $Y^1(z)$ then there is no reason for favoring a particular sign for $\pm Y^1(z)$. The conjugate interaction $\Phi ^*_{aux}(z)$ would be ``the other half'' of the sine-Liouville interaction and differs from $\Phi _{aux}(z)$ only in the chirality of the $U(1)$ charge $i\sqrt {\frac k2}\partial X^1(z)$. Moreover, such a term would actually be required in order to allow for the violation of the winding number conservation of the coset theory in a positive amount $\sum _{i=1}^N \omega _i >0$. Considering both terms together in the action leads to propose the correspondence between the WZNW model and the sine-Liouville ($\times \ time$) CFT through (\ref{rt}) since we could write
\begin{equation}
\Phi _{aux}(z) + \Phi ^* _{aux}(z) = 2e^{-\sqrt {\frac {k-2}{2}}\rho (z)} \cos \left( \sqrt {k/2} (Y_L ^1 (z)+Y_R ^1(\bar z))\right)    \label{psssi}
\end{equation}
where, again, $Y_L ^1(z)$ and $Y_R ^1(\bar z )$ refer to the left and right modes of the field respectively (cf. (\ref{taquion})). Besides, notice that if both $\Phi _{aux}^*(z)$ and $\Phi _{aux}(z)$ are included in the correlators, then the ``cosmological term'' $e^{\sqrt {2}b\varphi (z)}$ is not required anymore and can be excluded without restricting the generic correlators ({\it i.e.} if an analytic extension in the amount of screening charges $M_{\pm }$ is considered; which is usual in the Coulomb gas-type realization, cf. \cite{BB,GN3}). This is because, in that case, the conservations laws coming from the integration over the zero-modes of the fields would be the following ones
\begin{eqnarray}
&&\sum _{i=1}^N j _i +N-1+\frac {k-2}{2}(M_-+M_-)=0 \ \ \ (s=0) \\
&&\sum _{i=1}^N(m_i+\frac k2 \omega _i) = \sum _{i=1}^N(\bar {m}_i+\frac k2 \omega _i) = 0 \\
&&\sum _{i=1}^N \omega _i = (M_- - M_+)+2-N\ , \label{conservationII}
\end{eqnarray}
where $M_{+ }$ and $M_-$ refer to the amount of screening operators of the kind $\Phi _{aux}(z)$ and $\Phi ^*_{aux}(z)$ to be included when realizing the correlators following the Coulomb gas-like prescription; (cf. \cite{FH} after replacing the notation as $j _i \to -1-j_i$).

As it is well known, a similar relation between WZNW and sine-Liouville conformal models was originally conjectured by Fateev, Zamolodchikov and Zamolodchikov in their quoted unpublished paper, and this is the reason because it is known as the FZZ duality (see \cite{FB} for interesting discussions on related subjects). In the context we discussed here, a relation between the non-linear $\sigma $-model on $AdS_3$ and a sine-Liouville action seems to arise by a constructive procedure which simply follows from the identity (\ref{rt}), rigorously proven in \cite{RT,R}. It is worth emphasizing that, although it seems to be natural from the viewpoint of the symmetry under $\omega \to -\omega$, the inclusion of the term $\Phi ^*_{aux} (z)$ ({\it i.e.} the ``other half'' of sine-Liouville) in the action (\ref{S}) has to be assumed in order to eventually state such connection with the sine-Liouville field theory. However, the fact that both $\Phi _{aux}(z)$ and $\Phi ^*_{aux}(z)$ contain the appropriate exponents turns out to be a suggestive fact. Hence, within this framework, the sine-Liouville field theory can be seen to arise as a relatively natural free field realization of the associated WZNW theory. In the following paragraphs we will analyze the correlation functions in sine-Liouville theory in order to emphasize the similarities existing without free field realization (\ref{fggg}).


Let us begin by considering sine-Liouville model coupled to
matter. The action of this CFT is 
\begin{equation}
S=\frac{1}{2\pi }\int d^{2}z\left( \left( \partial \rho \right)
^{2}-\frac{R\rho }{\sqrt{2(k-2)}}+2 \ 
e^{-\sqrt{\frac{k-2}{2}}\rho }\cos \left( \sqrt{k/2}( Y^1
_{L}-Y^1 _{R}) \right) +\partial Y^{\mu }\bar{\partial}Y_{\mu }\right)
\end{equation}
Here, the matter sector is represented by the bosons $Y^{\mu }(z)$, where $\mu \in \{0,1,2,...d\}$. In Ref. \cite{FH} Fukuda and Hosomichi
proposed a free field realization of the correlation functions in this
theory by means of the insertion of two different screening operators\footnote{The screening operators realize the interaction term of sine-Liouville
action; see (\ref{taquion}).} and by evaluating a Dotsenko-Fateev type
integral representation  (we describe this representation in detail below). The three-point correlators representing processes violating winding number were explicitly computed by the authors; this was
achieved by the insertion of a different number of both screening charges.
Similar integral representations have been discussed in the free field realization of string theory on Euclidean
$AdS_{3}$ and in the case of two-dimensional black hole background. However, several
differences can be noticed if a detailed comparison of these models is
performed; the role played by the screening charges and the conjugated
representations of the identity operator are examples of these distinctions. 

The vertex operators creating primary states from the vacuum of the theory
can be written in terms of the bosonic fields $\rho (z)$ and $Y^{\mu }(z) $ as follows\footnote{Notice that we are indistinctly using the notation $\Phi $ to refer to both WZNW and sine-Liouville primaries.}
\begin{equation}
\Phi _{j,m,\bar{m};q_{i}}(z)=e^{j\sqrt{\frac{2}{k-2}}\rho (z)+i\sqrt{\frac{2}{k}}\left( mY^1 _{L}(z)-\bar{m}Y^1 _{R}(\bar{z} 
)\right) }e^{iq_{i}Y^{i}(z)} \label{jkl}
\end{equation}
with $i \in \{ 0, 2, 3, 4, ... d\}$. As before, the correlators of the fields are those of $U(1)$ bosons in two-dimensions; namely 
\begin{eqnarray}
\left\langle \rho (z_{a})\rho (z_{b})\right\rangle
=-2\log \left| z_{a}-z_{b}\right| \ , \ \ \left\langle Y^{\mu } (z_{a})Y^{\nu } (z_{b})\right\rangle =-2\eta ^{\mu \nu }\log \left| z_{a}-z_{b}\right| 
\end{eqnarray}
with $\eta ^{\mu \nu }=diag \{ -1, +1, +1, ... +1\}$ and $\mu , \nu \in \{ 0,1,2, ... d \}$. The conformal dimension of (\ref{jkl}) takes the form
\begin{eqnarray}
h _{(j,m,q)} &=&-\frac{j(j+1)}{k-2}+\frac{m^{2}}{k}+\frac{q_{i}q^{i}}{2}
\end{eqnarray}
satisfying the Virasoro constraint $h _{(j,m,q)}=\bar{h}_{(j,\bar{m 
},q)}=1$ as requirement for the string theory. The quantum numbers
parameterizing the spectrum fall within a lattice in the
following form 
\begin{eqnarray}
p &= &m+\bar{m}\in Z \\
\omega &= &\frac {m-\bar{m}}{k}\in Z
\end{eqnarray}
and the central charge of the model is given by 
\begin{equation}
c=d+2+\frac{6}{k-2}   \label{acordate2}
\end{equation}
Restriction $c=26$ implies $k=2(27-d)/(24-d)$ and is required to
define the string theory as well.

Now, let us move to the correlation functions and the factorization procedure. As a first step in our analysis, it is convenient to start by reviewing
the integral construction of Ref. \cite{FH} where a Coulomb gas-type realization was proposed. Actually, the treatment of the
sine-Liouville interaction term as a perturbation is as usual in the Feigin-Fuchs realization. The introduction of the interaction
effects can be viewed as the insertion of screening charges into the
correlation functions. In this case, the screening operators are given by 
\begin{equation}
\Phi _{\pm }(z)=\Phi _{1-\frac k2, \pm \frac k2, \pm \frac k2, 0}(z)=e^{\pm i\sqrt{\frac{k}{2}}\left(
Y^1 _{L}(z)-Y^1 _{R}(\bar{z})\right) -\sqrt{\frac{k-2}{2}}\rho (z)}  \label{screening}
\end{equation}
Then, the interaction term can be written as in (\ref{taquion}); namely 
\begin{equation}
\frac{1}{4\pi }\int d^{2}z\left( \Phi _{+} (z)+\Phi _{-} (z) \right)
\end{equation}
Roughly speaking, $\Phi _{+}(z)$ and $\Phi _{-}(z)$ play the role of $\Phi_{aux}(z)$ and respectively $\Phi _{aux}^* (z)$ in (\ref{psssi}). Accordingly, the Coulomb-gas like prescription, applied to the case of sine-Liouville theory, implies that the generic $N$-point function turns out to be proportional to 
\begin{eqnarray}
\sim \prod_{a=1}^{N}\int d^{2}z_{a}\prod_{r=1}^{s_{+}}\int d^{2}u_{r}\prod_{l=1}^{s_{-}}\int
d^{2}v_{l}  \left\langle \prod_{a=1}^{N}\Phi _{j_{a}m_{a},\bar{m} 
_{a},q_{a}}(z_{a})\prod_{r=1}^{s_{+}}\Phi _{+}(u_{r}) 
\prod_{l=1}^{s_{-}}\Phi _{-}(v_{l})\right\rangle . \label{aa}
\end{eqnarray}
which is the standard representation for the perturbative analysis of
conformal models, where the expectation value is defined with respect to the free theory ({\it i.e.} without the sine-Liouville interaction term since this is already taken into account due to the presence of the screenings $\Phi _{\pm}$). Then, by expanding the correlator, we find that this goes like
\begin{eqnarray*}
\prod_{a=1}^{N}\int
d^{2}z_{a}\prod_{a<b}^{N-1,N}\left| z_{a}-z_{b}\right| ^{-\frac{4j_{a}j_{b}}{ 
k-2}+2q_{a}^{i}q_{bi}} \left( z_{a}-z_{b}\right) ^{\frac{2}{k}m_{a}m_{b}}\left( \bar{z} 
_{a}-\bar{z}_{b}\right) ^{\frac{2}{k}\bar{m}_{a}\bar{m}_{b}}\times 
\end{eqnarray*}
\begin{eqnarray}
&&\ \times \prod_{r=1}^{s_{+}}\int d^{2}u_{r}\prod_{l=1}^{s_{-}}\int
d^{2}v_{l}\prod_{a=1}^{N}\left( \prod_{r=1}^{s_{+}}\left| z_{a}-u_{r}\right|
^{2(j_{a}+m_{a})}\left( \bar{z}_{a}-\bar{u}_{r}\right)
^{-p_{a}}\prod_{l=1}^{s_{-}}\left| z_{a}-v_{l}\right|
^{2(j_{a}-m_{a})}\left( \bar{z}_{a}-\bar{v}_{l}\right) ^{p_{a}}\right)
\times   \nonumber  \label{a} \\
&&\ \times \left( \prod_{r<t}^{s_{+}-1,s_{+}}\left| u_{r}-u_{t}\right|
^{2}\prod_{l<t}^{s_{-}-1,s_{-}}\left| v_{t}-v_{s}\right|
^{2}\prod_{l=1}^{s_{-}}\prod_{r=1}^{s_{+}}\left| v_{l}-u_{r}\right|
^{2-2k}\right)   \label{a}
\end{eqnarray}
The charge symmetry conditions, yielding from the integration over the
zero-modes of the fields, take in this case the form 
\begin{eqnarray}
\sum_{a=1}^{N}j_{a}+1 &=&\frac{k-2}{2}\left( s_{+}+s_{-}\right)  \label{ahora} \\
\sum_{a=1}^{N}m_{a} &=&\frac{k}{2}\left( s_{+}-s_{-}\right)  \\
\sum_{a=1}^{N}\bar{m}_{a} &=&\frac{k}{2}\left( s_{-}-s_{+}\right)  \\
\sum_{a=1}^{N}q_{a}^i &=&0.
\end{eqnarray}
It is instructive comparing (\ref{ahora}) with (\ref{antes}) (after performing the Weyl reflection $j\to -1-j$). This shows the parallelism between our realization of WZNW correlators and those quantities in sine-Liouville model.

Now, we are ready to see how the physical state conditions for the intermediate
string states arise from the pole structure of the $N$-point functions in
the factorization limit (in a similar way as we did it in the previous subsection for the WZNW correlators). In order to realize the factorization, we take
again the coincidence limit for the inserting points of a pair of vertex operators, 
{\it v.g.} $z_{1}\rightarrow z_{2}$. As we did it before, we take the limit of $\tilde{s}$ screening operators going to $z_{1}$ (being 
$0\leq \tilde{s}\leq s$), $i.e.$ by splitting the set of screening operators in two
different sectors of integration: those whose inserting points are taken as
going to $z_{1}$ and those whose inserting points remain fixed in the
worldsheet. Then, in order to take this limit conveniently, let us define the
following change of variables 
\begin{eqnarray}
\varepsilon &\equiv &z_{1}-z_{2}  \label{var1} \\
\varepsilon x_{r} &\equiv &z_{1}-u_{r}\quad ,\quad r\leq \tilde{s} 
_{+} \\
\varepsilon y_{l} &\equiv &z_{1}-v_{l}\quad ,\quad l\leq \tilde{s} 
_{-}  \label{var3}
\end{eqnarray}
By replacing this in expression (\ref{a}), we obtain that the (integrated) $N$-point correlators turns out to be proportional to
\begin{eqnarray}
&& \prod_{a=1}^{N}\int d^{2}z_{a}\prod_{a<b}^{N-1,N}\left|
z_{a}-z_{b}\right| ^{-\frac{4j_{a}j_{b}}{k-2}+\frac{4}{k} 
m_{a}m_{b}+2q_{a}^{i}q_{bi}}\left( \bar{z}_{a}-\bar{z}_{b}\right) ^{\frac{2}{ 
k}\left( \bar{m}_{a}\bar{m}_{b}-m_{a}m_{b}\right) }\times  \nonumber \\
&& \times \sum_{\tilde{s}_{+}=0}^{s_{+}}\sum_{\tilde{s}_{-}=0}^{s_{-}} 
\left| \varepsilon \right| ^{\tilde{s}_{+}(\tilde{s}_{+}+1)+\tilde{s}_{-}( 
\tilde{s}_{-}+1)+(2-2k)\tilde{s}_{+}\tilde{s}_{-}} C(s_+ , \tilde {s}_+ ) C(s_- , \tilde{s} _-) \times \nonumber \\
&& \times \prod_{r=1}^{\tilde{s}_{+}}\int d^{2}x_{r}\prod_{l=1}^{\tilde{s}_{-}}\int d^{2}y_{l}\prod_{r=\tilde{s}_{+}+1}^{s_{+}}\int d^{2}u_{r}\prod_{l= 
\tilde{s}_{-}+1}^{s_{-}}\int d^{2}v_{l}\times \left| W\left( \varepsilon
;z_{a},u_{r},v_{l}\right) \right| ^{2}\times  \nonumber \\
&& \times \prod_{a=1}^{n}\left( \prod_{r=\tilde{s}_{+}+1}^{s_{+}}\left|
z_{a}-u_{r}\right| ^{2(j_{a}+m_{a})}\left( \bar{z}_{a}-\bar{u}_{r}\right)
^{-p_{a}}\prod_{l=\tilde{s}_{-}+1}^{s_{-}}\left| z_{a}-v_{l}\right|
^{2(j_{a}-m_{a})}\left( \bar{z}_{a}-\bar{v}_{l}\right) ^{p_{a}}\right) \times
\nonumber \\
&& \times \prod_{r<t}^{\tilde{s}_{+}-1,\tilde{s}_{+}}\left|
x_{r}-x_{t}\right| ^{2}\prod_{l<t}^{\tilde{s}_{-}-1,\tilde{s}_{-}}\left|
y_{l}-y_{t}\right| ^{2}\prod_{l=1}^{\tilde{s}_{-}}\prod_{r=1}^{\tilde{s} 
_{+}}\left| y_{l}-x_{r}\right| ^{2-2k}\times  \nonumber \\
&& \times \prod_{t>r=\tilde{s}_{+}+1}^{s_{+},s_{+}-1}\left|
u_{t}-u_{r}\right| ^{2}\prod_{t>l=\tilde{s}_{-}+1}^{s_{-},s_{-}-1}\left|
v_{t}-v_{l}\right| ^{2}\prod_{l=\tilde{s}_{-}+1}^{s_{-}}\prod_{r=\tilde{s} 
_{+}+1}^{s_{+}}\left| v_{l}-u_{r}\right| ^{2-2k}  \label{b}
\end{eqnarray}
where we defined the function
\[
\left| W\left( \varepsilon ;z_{a},u_{r},v_{l}\right) \right|
^{2}=\prod_{a=1}^{N}\prod_{r=1}^{\tilde{s}_{+}}\left|
z_{a}-z_{1}+\varepsilon x_{r}\right| ^{2(j_{a}+m_{a})}\left( \bar{z}_{a}- 
\bar{z}_{1}+\bar{\varepsilon}\bar{x}_{r}\right) ^{-p_{a}}\times 
\]
\begin{eqnarray}
&&\times \prod_{l=1}^{\tilde{s}_{-}}\left| z_{a}-z_{1}+\varepsilon
y_{l}\right| ^{2(j_{a}-m_{a})}\left( \bar{z}_{a}-\bar{z}_{1}+\bar{\varepsilon 
}\bar{y}_{l}\right) ^{p_{a}}\times  \nonumber \\
&&\ \ \times \prod_{r=1}^{\tilde{s}_{+}}\prod_{t=\tilde{s} 
_{+}+1}^{s_{+}}\left| z_{1}-\varepsilon x_{r}-u_{t}\right| ^{2}\prod_{l=1}^{ 
\tilde{s}_{-}}\prod_{t=\tilde{s}_{-}+1}^{s_{-}}\left| z_{1}-\varepsilon
y_{l}-v_{t}\right| ^{2}\times  \nonumber \\
&&\ \ \times \prod_{l=\tilde{s}_{+}+1}^{s_{-}}\prod_{r=1}^{\tilde{s} 
_{+}}\left| z_{1}-v_{l}-\varepsilon x_{r}\right| ^{2-2k}\prod_{l=1}^{\tilde{s 
}_{-}}\prod_{r=\tilde{s}_{+}+1}^{s_{+}}\left| z_{1}-\varepsilon
y_{l}-u_{r}\right| ^{2-2k}  \label{c}
\end{eqnarray}
and where the sums over $\tilde{s}_{\pm }$ and the multiplicity factor $C(s_{\pm }, \tilde {s}_{\pm }) = \frac {\Gamma (s_{\pm}+1)}{\Gamma (\tilde {s}_{\pm }+1) \Gamma (s_{\pm }-\tilde{s}_{\pm }+1)}$ stand by considering all the different ways of selecting 
$\tilde{s}_{\pm }$ among $s_{\pm }$ screening operators $\Phi _{\pm } (z)$ to be
taken as going to $z_{1}$ in the factorization limit. As it can be seen directly from the definition of the variables $x$ and $y$ in (\ref{var1})-(\ref{var3}), the factorization limit is realized by
taking the limit $\varepsilon \rightarrow 0$. In order to explicitly
identify the sequence of poles that appears in this limit it will be useful to
expand the above expression in terms of powers of $\varepsilon $, namely 
\begin{equation}
\left| W\left( \varepsilon ;z_{a},u_{r},v_{l}\right) \right| ^{2}\equiv
W\left( \varepsilon ;z_{a},u_{r},v_{l}\right) \bar{W}\left( \bar{ 
\varepsilon};\bar{z}_{a},\bar{u}_{r},\bar{v}_{l}\right) =\sum_{n,\bar{n}\in 
{\bf N}^{2}}\left| W_{n,\bar{n}}\left( z_{a},u_{r},v_{l}\right) \right|
^{2}\cdot \varepsilon ^{n} \bar{\varepsilon}^{\bar{n}}
\end{equation}
where 
\begin{equation}
\left| W_{n,\bar{n}}\left( z_{a},u_{r},v_{l}\right) \right| ^{2}=\frac{ 
\partial _{\varepsilon }^{n}\partial _{\bar{\varepsilon}}^{\bar{n}}\left|
W\left( \varepsilon ;z_{a},u_{r},v_{l}\right) \right| _{(\varepsilon =\bar{ 
\varepsilon}=0)}^{2}}{\Gamma (n+1)\Gamma (\bar{n}+1)}  \label{taylor}
\end{equation}
Then, we can collect the whole dependence of $\left( \varepsilon ,\bar{ 
\varepsilon}\right) $ and therefore write down the following expression for
the power expansion in the neighborhood of $\varepsilon \simeq 0$ and write
\[
\int d^{2}\varepsilon \sum_{n,\bar{n}\in {\bf N}^{2}}\left| W_{n,\bar{n} 
}\left( z_{a},u_{r},v_{l}\right) \right| ^{2}\varepsilon ^{\eta
_{(j_{a},m_{a})}+n}\bar{\varepsilon}^{\eta _{(j_{a},-\bar{m}_{a})}+\bar{n}} 
\]
where we have denoted 
\begin{eqnarray*}
\eta _{(j_{a},m_{a})} &=&\left( \tilde{s}_{+}+\tilde{s}_{-}\right) \left(
j_{1}+j_{2}\right) +\frac{1}{2}\left( \tilde{s}_{+}+\tilde{s}_{-}\right)
\left( \tilde{s}_{+}+\tilde{s}_{-}+1\right) + \\
&&-k\tilde{s}_{+}\tilde{s}_{-}-2\frac{j_{1}j_{2}}{k-2}+2\frac{m_{1}m_{2}}{k} 
+\left( m_{1}+m_{2}\right) \left( \tilde{s}_{+}-\tilde{s}_{-}\right)
+q_{1}^{i}q_{2i}
\end{eqnarray*}
Hence, the pole condition is $\eta _{(j_{a},m_{a})}+n=\eta _{(j_{a},-\bar{m} 
_{a})}+\bar{n}=-1$ and can be written as follows 
\[
-\frac{\left( j_{1}+j_{2}-\left( \frac{k}{2}-1\right) \left( \tilde{s}_{+}+ 
\tilde{s}_{-}\right) \right) \left( j_{1}+j_{2}-\left( \frac{k}{2}-1\right)
\left( \tilde{s}_{+}+\tilde{s}_{-}\right) +1\right) }{k-2}+\frac{\left(
q_{1}^{i}+q_{2}^{i}\right) \left( q_{1i}+q_{2i}\right) }{2}+ 
\]
\[
+\frac{\left( m_{1}+m_{2}+\frac{k}{2}\left( \tilde{s}_{+}-\tilde{s} 
_{-}\right) \right) ^{2}}{k}-1=0, 
\]
and analogously for $\left( m_{a},\tilde{s}_{\pm }\right) \leftrightarrow
\left( \bar{m}_{a},\tilde{s}_{\mp }\right) $. Besides, if the physical constraint $h _{(j_{a},m_{a},q_{a})}=\bar{h}_{(j_{a}, \bar{m}_{a},q_{a})}=1$ is assumed to hold, the condition above can be written as a Virasoro constraint as well
\begin{eqnarray}
h _{(j,m,q)} &=&-\frac{j(j+1)}{k-2}+\frac{m^{2}}{k}+\frac{q^{i}q_{i}}{2} 
=1
\end{eqnarray}
and analogously for $\bar{h}_{(j,\bar{m},q)}$, where the quantum numbers $j,$ $p=m+\bar{m}$ and $\omega =(m-\bar{m})/k$ are
given by 
\begin{eqnarray}
j &=&j_{1}+j_{2}-\frac{k-2}{2}\left( \tilde{s}_{+}+\tilde{s}_{-}\right)
\label{r1} \\
p &=&p_{1}+p_{2} \ , \ \  q^{i} =q_{1}^{i}+q_{2}^{i}  \\
\omega &=&\omega _{1}+\omega _{2}+\tilde{s}_{+}-\tilde{s}_{-}
\end{eqnarray}
with $0\leq \tilde{s}_{\pm }\leq s_{\pm }$. Consequently, the following
identifications hold 
\begin{eqnarray}
\sum_{a=3}^{N}j_{a}+j+1 &=&\frac{k-2}{2}\left( s_{+}+s_{-}-\tilde{s}_{+}- 
\tilde{s}_{-}\right)  \label{r6} \\
\sum_{a=3}^{N}p_{a}+p &=&0 \ , \ \  \sum_{a=3}^{N}q_{a}^{i}+q^{i} =0  \\
\sum_{a=3}^{N}\omega _{a}+\omega &=&s_{-}-\tilde{s}_{-}+\tilde{s}_{+}-s_{+} \ .
\label{r7}
\end{eqnarray}
with $i\in \{ 0,2,3,4,...d\} $.

Hence, in this way, we have obtained the quantum numbers of the intermediate
channels in the tree-level scattering processes in sine-Liouville string
theory. Again, it was achieved by writing the pole condition arising in the
factorization limit as a Virasoro constraint for these physical states.

An interesting comment on the functional form of the
correlators in the coincidence limit is related to the factors involved in
the different terms of the power expansion in $\varepsilon $. The
characteristic property of the realization described here is the explicit
form in which the different contributions of internal states can be
recognized. Notice that in the limit $\varepsilon \rightarrow 0,$ the following factors
arise in the expression (\ref{b}) 
\begin{eqnarray}
&&\ \ \ \ \ \ \prod_{t=\tilde{s}_{+}+1}^{s_{+}}\left( \left(
z_{1}-u_{t}\right) ^{j_{1}+j_{2}+m_{1}+m_{2}+\tilde{s}_{+}-\tilde{s}_{-}-k 
\tilde{s}_{-}}\left( \bar{z}_{1}-\bar{u}_{t}\right) ^{j_{1}+j_{2}-\bar{m} 
_{1}-\bar{m}_{2}+\tilde{s}_{+}-\tilde{s}_{-}-k\tilde{s}_{-}}\right) \times 
\nonumber \\
&&\ \ \ \ \ \ \times \prod_{t=\tilde{s}_{-}+1}^{s_{-}}\left( \left(
z_{1}-v_{t}\right) ^{j_{1}+j_{2}-m_{1}-m_{2}+\tilde{s}_{+}-\tilde{s}_{-}-k 
\tilde{s}_{+}}\left( \bar{z}_{1}-\bar{v}_{t}\right) ^{j_{1}+j_{2}+\bar{m} 
_{1}+\bar{m}_{2}+\tilde{s}_{+}-\tilde{s}_{-}-k\tilde{s}_{+}}\right) 
\nonumber
\end{eqnarray}
and, in an analogous way, 
\begin{eqnarray*}
&&\ \ \ \ \ \ \prod_{a=3}^{N}\left| z_{1}-z_{a}\right| ^{-4\frac{ 
j_{a}(j_{1}+j_{2})}{k-2}+2j_{a}(\tilde{s}_{+}+\tilde{s} 
_{-})+2q_{ai}(q_{1}^{i}+q_{2}^{i})}\left( z_{1}-z_{a}\right) ^{\frac{2}{k} 
m_{a}(m_{1}+m_{2})+m_{a}(\tilde{s}_{+}-\tilde{s}_{-})}\times \\
&&\ \ \ \ \ \times \left( \bar{z}_{1}-\bar{z}_{a}\right) ^{\frac{2}{k}\bar{m} 
_{a}(\bar{m}_{1}+\bar{m}_{2})-\bar{m}_{a}(\tilde{s}_{+}-\tilde{s}_{-})}
\end{eqnarray*}
In fact, these are precisely the contributions required to express the
original $N$-point functions in terms of the product of two different
correlators on the sphere, since these factors allow to recover the $n=\bar{n 
}=0$ contribution, which is simply interpreted as the product of two
correlators involving only tachyonic (non-excited) states. Moreover, once the higher
derivative terms arising in the power expansion (\ref{taylor}) are taken
into account, then the derivatives $\partial _{\varepsilon }$ acting on the
factors in (\ref{c}) generate contributions which reproduce the contractions
arising in the operator product expansions of the form $\partial
_{z_{1}}\left\langle \rho (z_{1})\rho (z_{a})\right\rangle \simeq
(z_{a}-z_{1})^{-1}$, {\it e.g.} 
\[
\partial _{\varepsilon }\prod_{a=2}^{N}\prod_{r=1}^{\tilde{s} 
_{+}}(z_{a}-z_{1}+\varepsilon x_{r})_{(\varepsilon
=0)}^{j_{a}+m_{a}}=\prod_{a=1}^{N}\prod_{r=1}^{\tilde{s} 
_{+}}(z_{a}-z_{1})^{j_{a}+m_{a}}\sum_{l=1}^{\tilde{s}_{+}}x_{l} 
\sum_{b=2}^{N}(j_{b}+m_{b})(z_{b}-z_{1})^{-1} 
\]
where, after evaluating $\varepsilon =0$, each step of the sum over the index $b$ must
be put in correspondence with each term arising in the operator product
expansion between an operator $\partial \rho (z_{1})$ and an exponential
field $e^{j\sqrt{\frac{2}{k-2}}\rho (z_{a})}$ (and consequently, this is
analogous for the contributions of the $Y^{\mu } $ fields). Then, we recover the higher order contributions interpreted as
different terms in the sum over excited intermediate states.

Let us make a brief remarks about the symmetry under the interchange $(m,\bar{m 
},s_{\pm })\leftrightarrow (\bar{m},m,s_{\mp })$. This invariance of
the whole expression reflects the fact that the dissipation of
the winding number in the two dimensional theory is characterized directly
by the relative amount $s_{+}$ and $s_{-}$ corresponding to both
screening operators $\Phi _{+}(z)$ and $\Phi _{-}(z)$ respectively; this is due to the chiral nature of the interaction term (\ref{screening}). This is because of the symmetry existing under $\omega \leftrightarrow -\omega $.

\subsection{On winding number conservation}

Before concluding, a word on the prescription for computing correlators in sine-Liouville. Actually, let us consider in detail the structure of the correlation functions
representing processes that violate the winding number conservation in this CFT. This computation, originally presented in
Ref. \cite{FH}, is similar to previous analysis done for the case of the WZNW
model, although some substantial differences do exist between calculations in both CFTs. Perhaps, the main difference to be emphasized regards the role played by the screening operators: While in sine-Liouville theory the violation of the winding number is produced by the presence of the sine-Liouville
interaction term itself (which is manifestly represented in the free field
realization by the insertion of a different amounts of both screening
operators $s_{+}$ and $s_{-}$), the violation of the winding number
in the $SL(2,R)_k$ WZNW model is more confusing and involves the inclusion of conjugate representations of the identity operator (which is often called the ``spectral flow operator'', see \cite{GN3,MO3,GL}). An interesting (curious) feature of the non-conservation of the winding in these CFTs is the existence of a bound for such violation. In particular, in Ref. \cite{MO3} Maldacena and Ooguri explained how to understand the upper bound for the violation of the total winding number in $SL(2,R)_k$ WZNW correlators as being related to the $\hat {sl(2)}_k$ symmetry of the theory (see appendix D of Ref. \cite{MO3}). Then, the question arises as to whether could be possible to conclude that, in analogous way, the upper bound for the violation of winding number in sine-Liouville theory turns out to be related to such symmetry as well. If this is indeed the case (and it seems to be due to the conjectured FZZ duality), then it should be feasible to verify that such a strange feature ({\it i.e.} the fact that the non-vanishing correlators satisfy $|\sum_{a=1}^N \omega _a|\leq N-2$) turns out to be related to the fact that the FZZ sine-Liouville model satisfies a very particular relation between the radius $R$ of the compact direction $Y^1(z)$ and the background charge of the field $\rho (z)$; namely $R= \sqrt {\frac k2}$, $Q_{\rho }= \frac {1}{\sqrt {k-2}}$. Presumably, there would be no reason for the upper bound of the violation of winding number to exist in the ``generic'' sine-Liouville field theory besides the FZZ dual radius $R = \sqrt {\frac k2}$. 

In order to verify this idea, let us make some remarks on Fukuda-Hosomichi representation of sine-Liouville correlators: In a very careful analysis of the integral representation, the authors of
\cite{FH} were able to show that it is feasible to translate the
integrals $\prod_{r,l}\int d^{2}u_{r}\int d^{2}v_{l}$ over the whole complex
plane into the product of contour integrals. Then, the integral
representation (\ref{aa}) turns out to be described by standard techniques
developed in the context of rational conformal
field theory. Such techniques were used and extended in \cite{FH} in
order to define a precise prescription to evaluate the
correlators by giving the formula for the contour integrals in the
particular case of sine-Liouville field theory. The first step in the calculation
was to decompose the $u_{r}$ complex variables (resp. $v_{l}$) into two
independent real parameters ({\it i.e.} the real and imaginary part of $ u_{r}$) which take values in the whole real line. Secondly, a Wick
rotation for the imaginary part of $(u_{r})$ was performed in order to introduce a
shifting parameter $\varepsilon $ which was subsequently used to elude the poles in $ 
z_{a}$. Then, the contours are taken in such a way that the poles at $v_r \to z_{a}$
are avoided by considering the alternative order with respect to this
inserting points. A detailed description for the prescription
can be found in the section 3 of the paper, where the authors refer to the quoted
works by Dotsenko and Fateev \cite{dot3,dot1,dot2}. In the computation for sine-Liouville, Fukuda and Hosomichi proved that, in the case of three-point function, the winding can be violated up to $|\sum _{a=1}^{N=3}\omega _a|\leq N-2=1$ and, presumably, this is the same for generic $N$. The key point in obtaining such a constraint is noticing that the integrand  that arises in the Coulomb gas-like prescription in sine-Liouville model contains contributions of the form (see the last line in (\ref{b}))
\begin{eqnarray}
\int d^2v_{r} d^2v_{t} |v_r-v_{t}|^2 ... \label{llll}
\end{eqnarray}
for $0\leq r , t \leq s_-$ and the same for the points $u_l$ with $0\leq l \leq s_+$, and where the dots ``$...$'' stand for ``other dependences'' on the inserting points $v_r$ and $u_l$; these points are those where the screenings of the kind $\Phi _- (u)$ and $\Phi _+(v)$ are respectively inserted. As explained in \cite{FH}, and raised at the level of ``lemma'', ``the integral vanishes for certain alignments of contours'' due to the fact that the exponent of $ |v_r-v_{t}|$ in (\ref{llll}) is $+2$. Conversely, in the case where such exponent is generic enough (let us say $2\rho $, following the notation of \cite{FH}), the ``integral has a phase ambiguity due to the multi-valuedness of $|v_r-v_{t}|^{2\rho }$ in the integrand''. Then, it is concluded that those integrals containing two contours of $v_r $ and $v_{t}$ just next to each other, then the integral vanishes. And this precisely happens for all the contributions of those correlators satisfying $|s_+ -s_-| = | \sum_{a=1}^N \omega _a | > N-2$. This led Fukuda and Hosomichi to prove that, for the three-point function, there are only three terms that contribute: one with $ \sum_{a=1}^3 \omega _a =1$, a second with $ \sum_{a=1}^3 \omega _a =-1$ and the conserving one, $ \sum_{a=1}^3 \omega _a =0$. 

Here, we can make two comments regarding this point: First, notice that an identical feature appears in our free field representation of WZNW correlators (\ref{fggg}) and, in general, for the $N$-point function as well; namely: due to the presence of $M\leq N-2$ additional screening operators $\Phi _{aux} (w)$ we get contributions like (\ref{llll}) in the integral representation (with $w_i$ instead of $v_r$) due to the OPE 
\begin{equation}
\Phi _{aux } (w_i) \Phi _{aux } (w_j) \sim |w_i - w_j|^{2} ... \label{mimics}
\end{equation}
This is exactly what happens in sine-Liouville theory and consequently explains the upper bound for the violation of winding number in our free field realization of (\ref{rt}) as well. 

The second comment we find interesting is that we can actually give an answer about ``how to relate the $\hat {sl(2)}_k$ symmetry of sine-Liouville at the dual FZZ radius $R\sim \sqrt{k/2}$ and the existence of the upper bound for the violation of the winding''. The relation precisely comes from the fact that the exponent $+2$ in (\ref{llll}) does correspond to the particular value of the radius\footnote{The author thanks J.M. Maldacena for a conversation about this point.}, leading to the very particular OPE
\begin{equation}
\Phi _{\pm } (v_r) \Phi _{\pm } (v_t) \sim |v_r - v_t|^{2 \rho _{k,R}} ... = |v_r - v_t|^{2} ... \ ,  \label{cq}
\end{equation}
which mimics (\ref{mimics}) because of $\rho _{k,\sqrt {k/2}}=1$. The fact of demanding $\Phi _{\pm }(z)$ as being $(1,1)$-operators is not enough for affirming that the exponent in (\ref{cq}) is to be $2\rho _{k,R}=2$; instead, this only holds for the ``dual radius'' $R=\sqrt {k/2}$ if the theory is requested to have the appropriate central charge (\ref{acordate2}).

This actually shows that the selection rule $ | \sum_{a=1}^N \omega _a | \leq N-2$ does correspond to a very particular point of the space of parameters of sine-Liouville field theory; {\it i.e.} the particular point where it turns out to be dual to the $SL(2,R)_k/U(1)$ WZNW model. We point out again that the same happens in our realization (\ref{fggg}). 

\section{Conclusions}

The Ribault formula (\ref{rt}), as well as the particular case $M=N-2$ studied
in Refs. \cite{S, RT},
turns out to be a useful tool to study correlation functions in the non-compact
WZNW model \cite{GN}. This is because such formula states a very
concise relation
between these observables and those of the Liouville field theory, which
is certainly a better understood CFT. Here, we have gone further by arguing
that, besides a mathematical coincidence between correlators of a pair of
CFTs, Ribault formula is actually suggesting the equivalence between the worldsheet
theory of $AdS_3$ winding strings and a tachyonic linear dilaton background
that
contains Liouville theory as a particular factor; see (\ref{barrita}). Moreover, since (\ref
{rt}) was rigorously proven independently of such a realization, this actually
does demonstrate that string theory in $AdS_3$ can be represented in such a way. The description of worldsheet dynamics of strings in curved space in terms of
flat linear dilaton backgrounds establishes an interesting relation between
curved and flat exact solutions of non-critical string theory. The discrete
light-cone Liouville description of $AdS_3$ string theory, as well as the
quoted FZZ duality between the 2D black hole and the sine-Liouville model, are
examples of such kind of description. According to what we discussed here, what Ribault formula is actually
stating is an equivalence between the non-linear $\sigma $-model on $AdS_3$
and the CFT on the (tachyonic) linear dilaton background described by (\ref{S}).
And this equivalence holds at the level of the $N$-point correlation
functions. The fact that this equivalence is reminiscent of the FZZ duality and makes contact with the discrete light-cone approach turns out to be one of the main observations of this note. We also showed the connection with the Wakimoto free field representation of $\hat {sl(2)}_k$ algebra.

By analyzing the prescription for computing correlation functions and the factorization limit of these observables, we showed that the $N$-point function factorizes in two pieces, each one preserving the total winding number up to $N_{1,2}-2$ units (with $N_1 +N_2 =N+2$). By comparing with the Coulomb gas-like realization for the case of sine-Liouville field theory, we argued that such upper bound turns out to be related to the $\hat{sl(2)}_k$ symmetry of the theory and this fact is clearly manifested in the integral representation of the correlators. The intermediate states in the four-point functions can be obtained by studying the pole structure in the factorization, and both states with winding number $\omega =0 $ and $\omega =\pm 1$ arise in the intermediate channels. This is consistent with previous results.  

Then, in this note we proposed a new representation of string theory in Lorentzian $AdS_3$ space and this explicitly realize the Stoyanovsky-Ribault-Teschner formula (\ref{rt}) in terms of free fields. Besides, this prepares the basis for the next step: the study of the matrix model formulation within a similar context.

\newpage

{\bf Note:} In Ref. \cite{R} it is asserted that Prof. V. Fateev has also considered a free
field realization of the formula (\ref{rt}) in an unpublished paper. The author
of this note would be interested in confirming the agreement between that
realization and the one presented here. In any case, the realization analyzed here presents more evidence in favour of the conjecture made in \cite{R} about the validity of the correspondence between WZNW and Liouville correlators for the maximally winding violating case (see subssection 3.2.2 of \cite{R}).
\[
\]
{\bf Note added:} After the first version of this paper appeared in arXives, the author received a copy of the unpublished work \cite{Fateev} from Prof. V. Fateev. There, a free field representation of the theory is presented and several results agree with those of the section 2 here. The author would like to thank Vladimir Fateev for kindly shareing his note.
\[
\]
{\bf Acknowledgement:} The author is grateful to Juan M. Maldacena for helpful comments and suggestions, and he thanks Ari Pakman and Yu Nakayama for several discussions on related topics as well. He would also like to express his gratitude to the Institute for Advanced Study, Princeton, for the hospitality during his stay, where the first part of this work was done.

This work was partially supported by Universidad de Buenos Aires and Instituto de F\'{\i}sica de La Plata, CONICET, Argentina.

\end{document}